\documentclass[10pt,a4paper]{article}

\usepackage[T1,T2A]{fontenc}
\usepackage[utf8]{inputenc}
\usepackage[russian, english]{babel}

\usepackage[affil -it]{authblk}

\usepackage{graphicx}
\usepackage{amsmath,amsfonts,amssymb,textcomp}
\usepackage{color}
\usepackage{xcolor}

\usepackage{amsmath}

\usepackage[sorting=none, style=phys, backend=biber, uniquename=init, giveninits=true]{biblatex}
\bibliography{references}

\usepackage[colorlinks]{hyperref}
\usepackage[colorinlistoftodos, textsize=tiny, textwidth=110,]{todonotes}
\setuptodonotes{fancyline}

\usepackage{array,multirow,tabularx, makecell}
\usepackage{adjustbox}

\usepackage[font=scriptsize]{caption}

\usepackage{sectsty}
\sectionfont{\centering}

\usepackage{indentfirst}

\title{\textbf{Interaction of solar neutrinos with $^{98,100}$Mo isotopes and the influence of nuclear resonances}}
\author[1,*]{ Yu.\,S.\, Lutostansky}
\author[1]{ N.\,A.\, Belogortseva}
\author[1,2,3]{A.\,N.\,Fazliakhmetov}
\author[1,2,**]{G.\,A.\, Koroteev}
\author[1]{ A.\,Yu.\, Lutostansky}
\author[1]{N.\,V.\, Klochkova}
\author[1]{A.\, P.\, Osipenko}
\author[1]{ V.\,N.\, Tikhonov}
\author[1]{ E.\,Yu.\, Zemskov}

\affil[1]{National Research Centre "Kurchatov Institute", Moscow, Russia}
\affil[2]{Moscow Institute of Physics and Technology (National Research University), Dolgoprudny, Russia }
\affil[3]{Institute for Nuclear Research of Russian Academy of Sciences, Moscow, Russia}

\affil[ ]{ }
\affil[ ]{E-mail:}
\affil[*]{ lutostansky@yandex.ru}
\affil[**]{ koroteev@phystech.edu}

\begin{document}

\maketitle

\begin{abstract} 
The process of neutrino interaction with $^{98}$Mo and $^{100}$Mo isotopes is studied taking into consideration the effect of charge-exchange resonances. The results obtained by calculating the cross section $\sigma(E_{\nu})$ for solar neutrino capture by the isotopes $^{98}$Mo and $^{100}$Mo are presented. Both the experimental data on the strength functions $S(E)$ obtained in charge-exchange reactions $(p, n)$ and $(^{3}\mathrm{He}, t)$ and the strength functions $S(E)$ calculated within the theory of finite Fermi systems were used. The effect of the resonance structure of $S(E)$ on the calculated cross sections for solar-neutrino capture is studied, and the contribution of each high-lying resonance to the capture cross section $\sigma(E_{\nu})$ is determined. The contributions of all components of the solar neutrino spectrum are calculated. The contribution of background solar neutrinos to the double-beta decay of $^{100}$Mo nuclei is estimated.
\end{abstract}

\section{INTRODUCTION}
In neutrino physics and astrophysics, the process of interaction of neutrino with matter is of great importance.
In most cases, it is needed to calculate neutrino capture cross-section $\sigma(E_{\nu})$ and take into account the structure of the charge-exchange strength function $S(E)$, which determines the magnitude of $\sigma(E_{\nu})$ and its energy dependence.

When calculating the cross sections of solar neutrinos interaction with atomic nuclei, $\sigma(E_{\nu})$, it is necessary to calculate the structure of the function $S(E)$ up to an energy of 19 -- 20~MeV. For solar neutrinos, the upper boundary of the spectrum is determined by the $hep$ reaction ${}^{3}\mathrm{He} + p \rightarrow {}^{4}\mathrm{He} + e^+ + \nu_e$, in which case $E_x \le 18.77$~MeV~\cite{Bahcall_book}. For the isotopes $^{98}$Mo and $^{100}$Mo under consideration, the strength functions $S(E)$ were measured up to $E_x = 18$~MeV for $^{98}$Mo~\cite{Rapaport_1985_98Mo} and in the region of $E_x < 20$~MeV for $^{100}$Mo~\cite{AKIMUNE199723},\cite{Thies_2012_100Mo}. The isotopes $^{98}$Mo and $^{100}$Mo differ in structure only by two neutrons, but, in the cross section for solar neutrino capture, $\sigma(E_{\nu})$, they differ many times, and this is what we will discuss in the present article.

Yet another reason why we have chosen these nuclei is that large-scale international projects aimed at studying double-beta decay employ the isotope $^{100}$Mo, and the influence of background solar neutrinos is very important. In the NEMO-3 experiment using 6.914~kg of the isotope $^{100}$Mo and 0.932~kg of the isotope $^{82}$Se, the half-life of $^{100}$Mo to the ground state of $^{100}$Ru was measured~\cite{Arnold_NEMO3}. When planning experiments with significantly higher exposure, the background from solar neutrinos cannot be neglected. These backgrounds will be taken into account for the SuperNEMO project involving a higher mass and a greater number of isotopes~\cite{RakhimovBarabash}. 
The situation around backgrounds is similar in the CUPID-Mo experiment, which is being performed at the Laboratoire Souterrain de Modane (LSM, France)~\cite{Armengaud_cupid-mo},
\cite{Augier_cupid-mo}, with promising development~\cite{CUPID-Alfonso2022} and at the initial stage of the AMoRE experiment~\cite{AMORE-1}, \cite{AMORE_Lee_2020}. 
For the $^{98}$Mo isotope, $\beta\beta$-decay is also energetically possible, and the possibility of this process is now beginning to be investigated~\cite{Nesterenko-98Mo}.

\begin{figure}[ht!]
	\centering
	\includegraphics[width=0.7\linewidth]{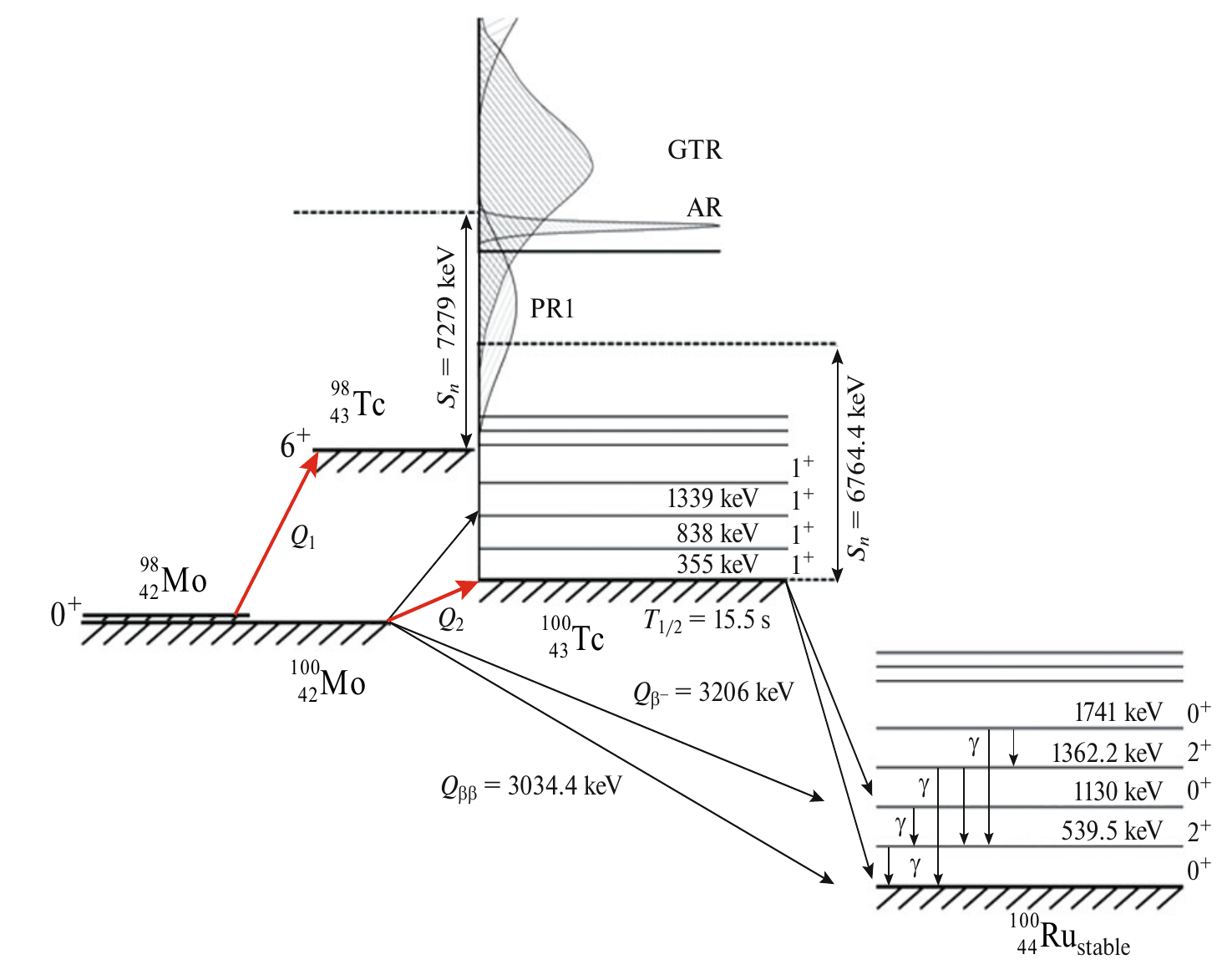}
	\caption{Scheme of charge-exchange excitations of $^{98, 100}$Mo nuclei.}
	\label{fig:1}
\end{figure}

Figure~\ref{fig:1} shows the scheme of charge-exchange excitations of $^{98, 100}$Mo nuclei upon neutrino capture followed by the decay of arising $^{98, 100}$Tc nuclei. One can see that the excited states of technetium isotopes have a resonance structure.

The giant Gamow–Teller (GT) resonance is the most intensive~\cite{Gaponov-Lutostansky-JETP-Lett}. An isobaric analog resonance (AR) lies below GTR~\cite{Gaponov-Lutostansky-Nucl-Phys}, while the so-called pygmy resonances (PR)~\cite{Lutostansky2017_JETP}, which are of importance in charge-exchange reactions~\cite{Pham-PhysRevC},~\cite{Lutostansky_Tikhonov2018} and in beta-decay processes~\cite{Verney-PhysRevC.95.054320}, lie still lower. 
Accordingly, these charge-exchange resonances manifest themselves in the strength function $S(E)$ and change substantially the results of the calculation of cross sections for charge-exchange reactions, including the cross sections $\sigma(E_{\nu})$ for solar neutrinos capture by atomic nuclei~\cite{Lutostansky_Tikhonov2018}, \cite{Lutostansky:2019iri}.

Figure~\ref{fig:1} also shows the energy thresholds $Q_1$ and $Q_2$ for neighboring $^{98}$Tc and $^{100}$Tc isotopes, respectively, which differ greatly. Thus, the energy $Q_1 = Q_{\beta}$  for the $^{98}$Tc isotope is $1684 \pm 3$~keV, and for $^{100}$Tc $Q_2 = 172.1 \pm 1.4$~keV~\cite{Huang_2021}. As a result, a dominant role in the process of solar-neutrino capture is played by hard solar neutrinos in the case of the $^{98}$Mo nucleus and by neutrinos of lower energy in the case of the $^{100}$Mo nucleus.

In the latter case, these are primarily $pp$ solar neutrinos (that is, those from the reaction $p + p \rightarrow {}^{2}H + e^+ + \nu_e$), with $E_x \le 420$~keV~\cite{Bahcall_book} and whose number is several orders of magnitude greater. This is the reason why the cross sections $\sigma(E_{\nu})$ for solar neutrinos capture by these nuclei differ strongly (see below).

\section{CHARGE-EXCHANGE EXCITATIONS OF THE $^{98, 100}$Mo ISOTOPES}

\begin{figure}[ht!]
	\centering
	\includegraphics[width=0.7\linewidth]{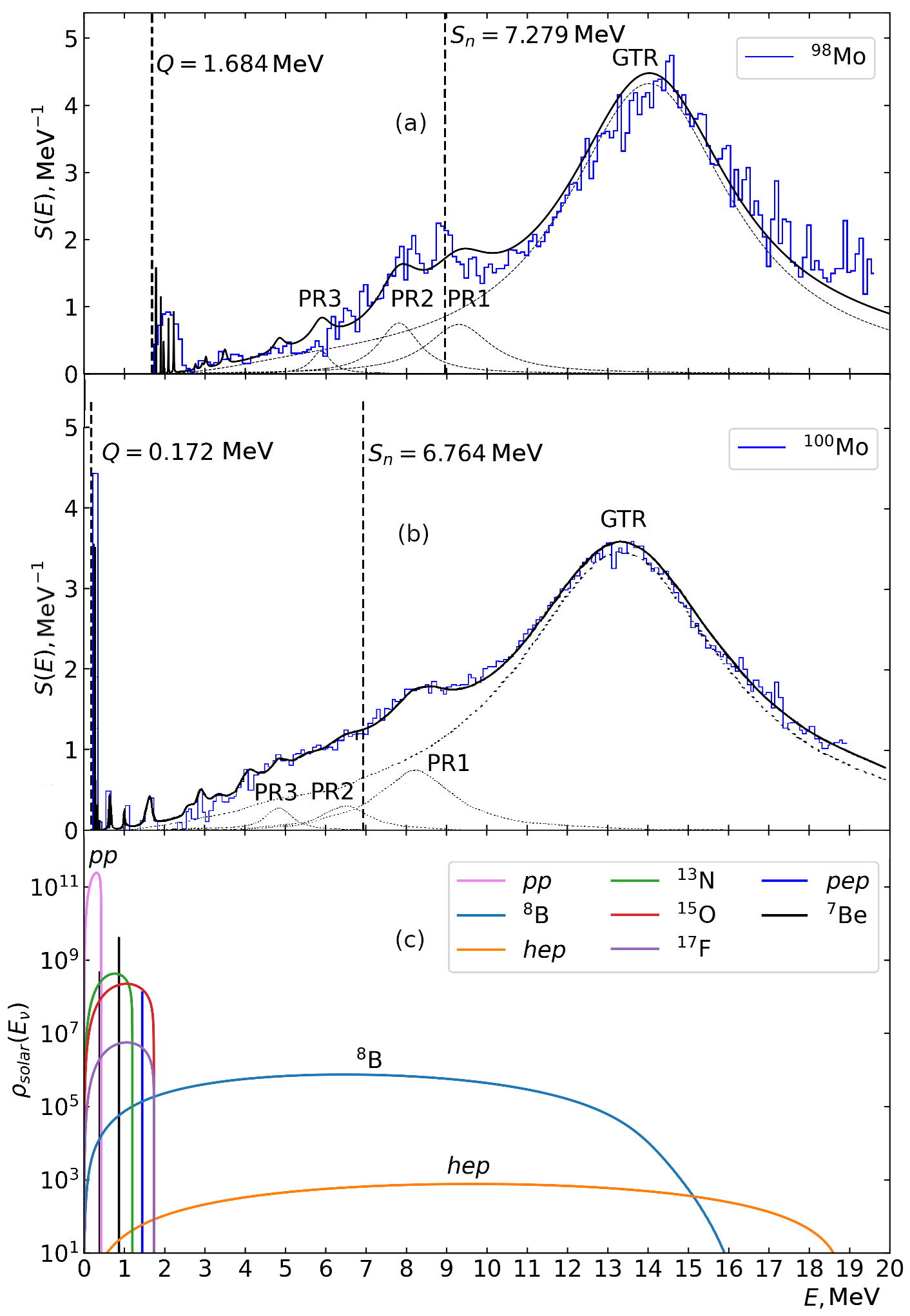}
	\caption{Scheme of charge-exchange excitations of $^{98, 100}$Mo nuclei. Charge-exchange strength function $S(E)$ for GT excitations of the isotopes \textit{(a)} $^{98}$Tc and \textit{(b)} $^{100}$Tc: (thin curves) experimental data from~\cite{Rapaport_1985_98Mo} for $^{98}$Tc and from~\cite{Thies_2012_100Mo} for $^{100}$Tc, (thick curves) results of the present calculation based on the theory of finite Fermi systems, and (dashed curves) resonances (GTR, PR1, PR2, and PR3). \textit{(c)} Fluxes of solar neutrinos, where various contributions are indicated.}
	\label{fig:2}
\end{figure}

The resonance structure of charge-exchange excitations of the $^{98, 100}$Mo nuclei is illustrated in Fig.~\ref{fig:2}, where the experimental data obtained for the strength functions in the reactions $^{98}\mathrm{Mo}(p, n)^{98}\mathrm{Tc}$~\cite{Rapaport_1985_98Mo} and $^{100}\mathrm{Mo}(^{3}\mathrm{He}, t)^{100}\mathrm{Tc}$ ~\cite{AKIMUNE199723},\cite{Thies_2012_100Mo} are shown along with the respective data calculated~\cite{Lutostansky2022} within the theory of finite Fermi systems (TFFS)~\cite{Migdal_book}.
The data in Fig.~\ref{fig:2} are given in the form of a graph that represents the dependence of the strength function $S(E)$ on the excitation energy $E$ reckoned from the ground state of the isotope $^{100}$Mo (see Fig.~\ref{fig:1}). 
Reckoned with respect to this reference value, the energies of the isobaric resonances have close values, since the isotopes $^{98}$Mo and $^{100}$Mo differ by only two neutrons. 
The same reference system permits determining the types of solar neutrinos in the graph (see Fig.~\ref{fig:2}\textit{b}) that make contributions in various regions of energies of the isotopes $^{98, 100}$Mo considered here. One can see that low-energy solar neutrinos (see Fig.~\ref{fig:2}\textit{c}) make a dominant contribution, which is several orders of magnitudes larger than the contribution of other neutrinos of the solar spectrum, to the capture cross section $\sigma(E_{\nu})$ for the $^{100}$Mo nucleus, but this is not so for $^{98}$Mo, in which case $Q_{\beta} = 1684$~keV and where a dominant contribution comes from harder boron and hep neutrinos (see Fig.~\ref{fig:2}\textit{c}).

The charge-exchange strength functions $S(E)$ for the isotopes $^{98, 100}$Mo were calculated within the microscopic theory of finite Fermi systems (TFFS)~\cite{Migdal_book} in just the same way as this was done earlier for other nuclei~\cite{Lutostansky2018_EPJ}, \cite{LUTOSTANSKY-Physics-Letters-B}. 
The energies of the excited states of the daughter nucleus and their matrix elements were determined by solving the set of secular TFFS equations for the effective field according to~\cite{Lutostansky2018_EPJ}, \cite{BORZOV1995335}. 
The calculations were performed in the coordinate representation with allowance for pairing in the single-particle basis.
The basis was taken in the Woods–Saxon model, and the subsequent iteration procedure was used to construct the nuclear potential. 
Effects of the change in the pairing gap in an external field were neglected—that is, it was assumed that $d^{1}_{pn} = d^{2}_{pn} = 0$. 
This is justified for an external field whose diagonal elements are zero (see~\cite{Migdal_book}, p. 200).

In this work, we used a simplified version of the 
research~\cite{BORZOV1995335} -- that is, a partial agreement 
with $m^{*} = m$ 
and with the local nucleon–nucleon interaction for allowed transitions $F_{\omega}$ in the Landau–Migdal form~\cite{Migdal_book}:
\begin{equation} 
	\label{eq:F_omega}
	F_{\omega} = C_{0}(f^{'}_{0} + g^{'}_{0} (\vec{\sigma_{1}} \vec{\sigma_{2}}))(\vec{\tau_{1}}\vec{\tau_{2}})\delta(\vec{r_{1}} - \vec{r_{2}})
\end{equation}
where $C_{0} = (\mathrm{d} \rho / \mathrm{d} \epsilon_{F})^{-1} = 300~$MeV$\cdot$\emph{fm}$^{3}$ ($\rho$~is the average nuclear-matter density) and $f^{'}_{0}$  and $g^{'}_{0}$ are the parameters of, respectively, isospin–isospin and spin–isospin quasiparticle interactions. These coupling constants are the parameters of the theory. In the present calculations, we used the values:
\begin{equation} 
	\begin{split}
		f^{'}_{0}= 1.351 \pm 0.027 \\	
		g^{'}_{0}= 1.214 \pm 0.048
	\end{split}
	\label{eq:constants}
\end{equation}
which were obtained recently~\cite{Lutostansky2020} from an analysis of the experimental data on the energies of the analog (38 nuclei) and GT (20 nuclei) resonances.
The energies, $E_i$, and the squares of the matrix elements, $M^2_i$, were calculated for allowed-transition-excited isobaric states of $^{98, 100}$Tc daughter nuclei. 
The continuous part of the spectra of the strength function $S(E)$ was calculated in the same way a~\cite{Lutostansky_Tikhonov2018} upon taking into account Breit–Wigner broadening (see~\cite{Lutostansky_Shulgina_PhysRevLett.67.430}).

For the isotope $^{98}$Tc, the charge-exchange strength function $S(E$) calculated for GT excitations of $^{98}$Mo is shown in Fig.~\ref{fig:2}\textit{a)}. The energies calculated for the GT resonance and for the 1, 2, and 3 pygmy resonances are the following: $E_{\mathrm{GTR}} = 12.45$~MeV, $E_{\mathrm{PR1}} = 7.32$~MeV, $E_{\mathrm{PR2}} = 6.10$~MeV, and $E_{\mathrm{PR3}} = 4.40$~MeV. The experimental value for the GT resonance is $E_{\mathrm{GTR}} \approx 12.3$~MeV~\cite{Rapaport_1985_98Mo} -- that is, the difference between the calculated and experimental values is moderately small and is equal to 0.15~MeV. 
As for the observed pygmy resonance, at $E_{\mathrm{PR}} = 6.78$~MeV according to the (B–W) fit, our calculations yield two (PR1 and PR2) closely lying resonances, and this looks like a fine structure of the observed PR peak. 
The results of our calculations show moderately weak resonances at the energies of 1.82 and 3.16~MeV; they correspond to small peaks observed in the vicinity of 2.1 and 3.4~MeV in the experiment reported in~\cite{Rapaport_1985_98Mo}. 
At low excitation energies of $E_x < 1$~MeV, the strength function calculated for $^{98}$Tc features several excited states that may correspond the observed low-lying excitations of $^{98}$Tc. The calculated value of the analog-resonance energy is $E_{\mathrm{AR}} = 9.78$~MeV, while its experimental counterpart is $E_{\mathrm{AR}} = 9.7$~MeV~\cite{Rapaport_1985_98Mo}. The difference is $\Delta E_{\mathrm{AR}} = 80$~keV and is comparable with the result of our previous calculation based on the TFFS approach, where $\Delta E_{\mathrm{AR}} = 110$~keV~\cite{Gaponov-Lutostansky-Nucl-Phys}.

For the isotope $^{100}$Tc, the charge-exchange strength function $S(E)$ calculated for GT excitations of the isotope $^{100}$Mo is presented in Fig.~\ref{fig:2}\textit{b)}. The calculated resonance energies are $E_{\mathrm{GTR}} = 13.20$~MeV, $E_{\mathrm{PR1}} = 8.09$~MeV, $E_{\mathrm{PR2}} = 6.32$~MeV, and $E_{\mathrm{PR3}} = 4.65$~MeV, and the experimental value for the GT resonance is $E_{\mathrm{GTR}} \approx 13.3$~MeV~\cite{AKIMUNE199723}; that is, the difference between the experimental and calculated values is small, 0.10~MeV. As for the observed pygmy resonance, at $E_{\mathrm{PR1}} = 8.0$~MeV~\cite{AKIMUNE199723} according to the (B–W) fit and $E_{\mathrm{PR1}} = 7.52$~MeV~\cite{Thies_2012_100Mo} according to the (G) fit, the calculated value turned out to be closer to the experimental value from~\cite{AKIMUNE199723} than to the experimental value from~\cite{Thies_2012_100Mo}. Low-lying excitations found in~\cite{AKIMUNE199723} (which was published earlier) at the energies of $E_1 = 1.4$ and $E_2 = 2.6$~MeV also appeared in the present calculations as the doublet of states at the energies of 1.30 and 1.42~MeV and an isobaric state at $E_2 = 2.70$~MeV.
The results calculated for the analog resonances are close to their experimental counterparts -- for example, the calculated value is $E_{\mathrm{AR}} = 10.99$~MeV, while the experimental one is $E_{\mathrm{AR}} = 11.085$~MeV~\cite{Thies_2012_100Mo}. The difference is $\Delta E_{\mathrm{AR}} = 95$~keV, and this is commensurate with the results of our earlier calculations~\cite{Gaponov-Lutostansky-Nucl-Phys}.

\section{NORMALIZATION OF STRENGTH FUNCTION AND \textit{QUENCHING}-EFFECT}

In describing both the experimental and the calculated data on the strength function $S(E)$ for the isotopes $^{98, 100}$Mo which are presented in Fig.~\ref{fig:2}, the normalization of $S(E)$ is an issue of importance. For example, the experimental data for $^{98}$Mo were obtained in the reaction  $^{98}\mathrm{Mo}(p, n)^{98}\mathrm{Tc}$~\cite{Rapaport_1985_98Mo}, whereupon the charge-exchange strength function $S(E)$ was determined up to the excitation energy of $E_{max} = 18$~MeV. 
It was found that the total sum of the GT matrix elements \textit{B}(GT) up to the energy of 18~MeV is $28 ± 5$~\cite{Rapaport_1985_98Mo}, which is $0.67 \pm 0.08$ of the maximum value of $3(N - Z) = 42$, which is given by the sum rule for GT excitations of the  $^{98}$Mo nucleus. 
This means that there is a deficit in the sum rule for GT excitations.

In the study~\cite{Thies_2012_100Mo}, the results of processing \textit{B}(GT) for  $^{100}$Mo are given over the energy range extending up to 4~MeV. For other energy values, the authors of the paper~\cite{Thies_2012_100Mo} do not present neither the dependence of \textit{B}(GT) on the energy $E$, nor do they give the sum $\sum B(\mathrm{GT})$. In the earlier study~\cite{AKIMUNE199723}, it was found, however, that the sum of GT matrix elements up to the energy of 18.8~MeV is 34.56 or 0.72 (72$\%$) of the maximum possible value of $3(N - Z) = 48$. This is greater by 7.5$\%$ than the respective result for $^{98}$Mo~\cite{Rapaport_1985_98Mo}.

The observed deficit in the sum rule for GT excitations is due to the \textit{quenching}-effect~\cite{ARIMA1999260} or to a violation of the normalization of GT matrix elements. Thus, according to the known sum rule, for GT-transitions, the normalization has the form:
\begin{equation} 
	\label{eq:Summ.M_i}
	\sum M^{2}_{i} = \sum B_i(\mathrm{GT}) = q[3(N-Z)] = e^{2}_{q}[3(N-Z)] \approx \int_{0}^{E_{max}} S(E) dE = I(E_{max})
\end{equation}
where $E_{max}$ is the maximum energy taken into account in the calculation or in the experiment and $S(E)$ is the charge-exchange strength function. 
In the present calculations, the value of $E_{max} = 20$~MeV was employed for the isotopes $^{98}$Mo and $^{100}$Mo, while, in the experiments, this energy was set to, respectively, $E_{max} = 18$~MeV~\cite{Rapaport_1985_98Mo} and $E_{max} \approx 19$~MeV~\cite{Thies_2012_100Mo}.
The parameter $q < 1$, in Eq.~(\ref{eq:Summ.M_i}) determines the $\textit{quenching}$-effect (deficit in the sum rule); at $q = 1$, $\sum M^{2}_{i} = \sum B_i(\mathrm{GT}) = 3(N - Z)$, which corresponds to the maximum value. Within the TFFS framework, $q = e^2_q$, where $e_q$ is an effective charge~\cite{Migdal_book}. 
As was shown by A. B. Migdal~\cite{Migdal-1957}, the effective charge should not exceed unity; 
for Fermi transitions, we have $e_q(F) = 1$, while, for GT transitions, $e_q(GT) = 1–2\zeta_S$ (see~\cite{Migdal_book}, p. 223]), where $\zeta_S$, $0 < \zeta_S < 1$, is an empirical parameter. 
Thus, we see that, in the case of Mo $\rightarrow$ Tc transitions considered here, the effective charge $e_q = e_{q}(\mathrm{GT})$ is a parameter that is extracted from experimental data. 

\begin{figure}[ht!]
	\centering
	\includegraphics[width=0.7\linewidth]{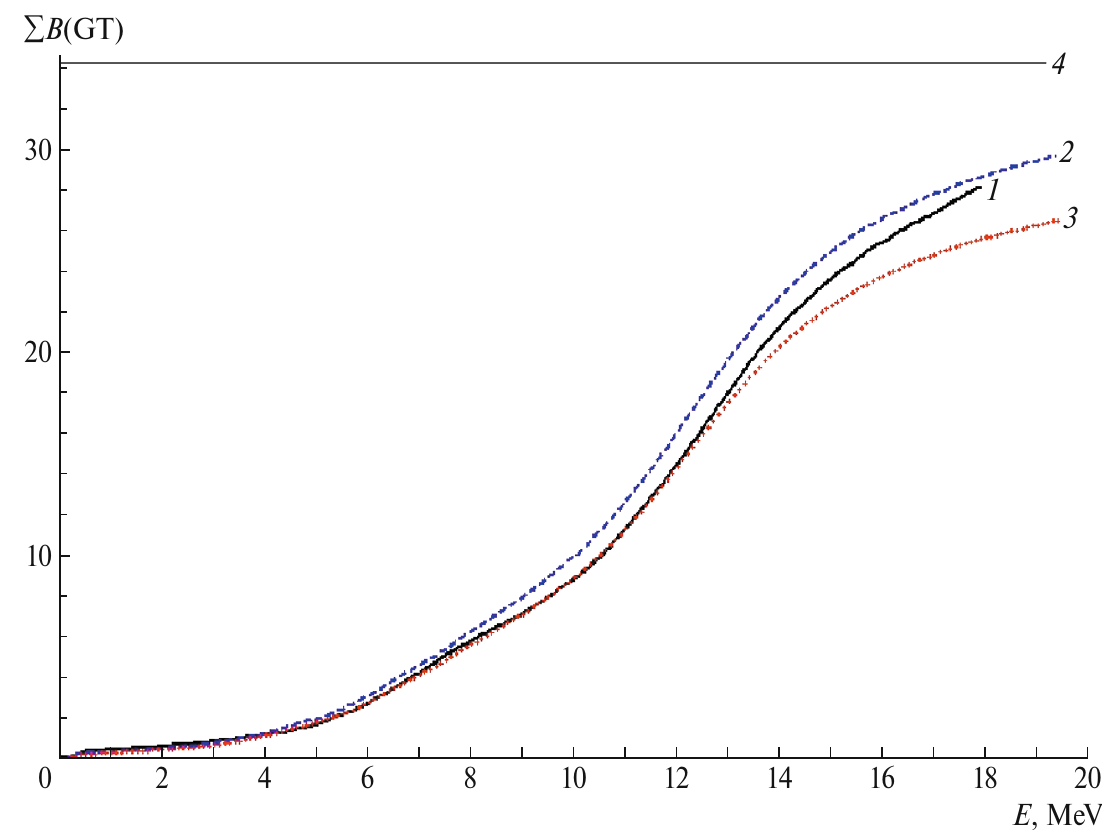}
	\caption{Sum  of the matrix elements for the isotope $^{98}$Mo as a function of the energy $E$ reckoned in $^{98}$Tc. The lines represent: \textit{(1)} experimental data from~\cite{Rapaport_1985_98Mo}, \textit{(2)} the results of our calculations based on the theory of finite Fermi systems at $e_q = 0.90$, \textit{(3)} the results of our calculations at $e_q = 0.80$, and \textit{(4)} the maximum value $\sum B_i(\mathrm{GT}) = q[3(N-Z)]$ at $q = e_q^2 = 0.9^2 = 0.81$.}
	\label{fig:3}
\end{figure}

Figure~\ref{fig:3} shows the sum $\sum B_i(\mathrm{GT})$ in (\ref{eq:Summ.M_i}) of the matrix elements for the isotope $^{98}$Mo as a function of the energy $E$ reckoned in the isotope $^{98}$Tc.
One can see that the calculations with the effective charge of $e_q = 0.9$ $(q = 0.81)$ describe the experimental data better, but, in the region of energies below 14~MeV, the calculations with the effective charge of $e_q = 0.8 (q = 0.64)$ lead to results that are closer to the experimental data. At higher energies, the calculated curve asymptotically tends to the value of $q[3(N - Z)] = q \cdot 42$ at $q = e_q^2 = 0.9^2 = 0.81$ (81$\%$).

\begin{figure}[ht!]
	\centering
	\includegraphics[width=0.7\linewidth]{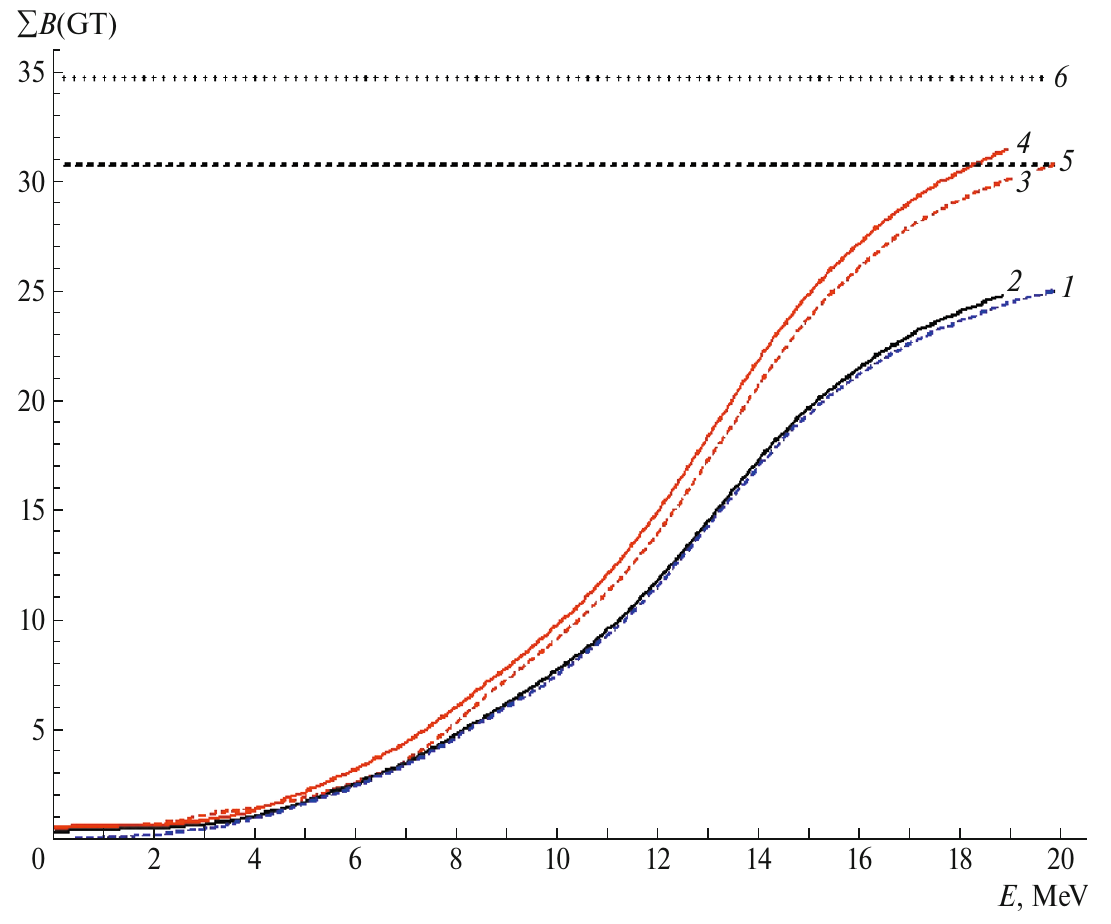}
	\caption{Sum  of the matrix elements for the isotope $^{100}$Mo as a function of the energy $E$ reckoned in the isotope $^{100}$Tc. The lines represent: \textit{(1)} the results of our calculations based on the theory of finite Fermi systems (TFFS) at $e_q = 0.80$, \textit{(2)} experimental data from~\cite{Thies_2012_100Mo},  \textit{(3)} the results of our TFFS calculations at $e_q = 0.85$, \textit{(4)} experimental data from~\cite{AKIMUNE199723}, \textit{(5)} the maximum value of the sum $\sum B_i(\mathrm{GT}) = q[3(N-Z)]$ at $q = e_q^2 = 0.80^2$, and \textit{(6)} the maximum value of the sum $\sum B_i(\mathrm{GT})$ at $q = e_q^2 = 0.85^2$.}
	\label{fig:4}
\end{figure}

Figure~\ref{fig:4} shows the sum $\sum B_i(\mathrm{GT})$ of matrix elements in (\ref{eq:Summ.M_i}) for the isotope $^{100}$Mo as a function of the energy $E$ reckoned in the isotope $^{100}$Tc.
For this isotope, the situation is more complicated than that in the case of the isotope $^{98}$Mo, because the available data from two experiments reported in~\cite{AKIMUNE199723} and~\cite{Thies_2012_100Mo} differ somewhat from each other in normalization of the strength function $S(E)$. In~\cite{AKIMUNE199723}, for example, the sum of GT matrix elements up to the energy of 18.8~MeV is 34.56 or $q = 0.72$, which is 72$\%$ of $3(N - Z) = 48$. This is 7.5$\%$ larger than that for $^{98}$Tc~\cite{Rapaport_1985_98Mo} and corresponds to the value of $e_q = 0.85$. In~\cite{Thies_2012_100Mo}, there are no data on the dependence of \textit{B}(GT) on $E_x$ for all of the energies presented there; nor does that article give the value of the $\sum B_i(\mathrm{GT})$. For the isotope $^{100}$Mo we have calculated the GT strength function $S(E)$ for two versions of normalization: $e_q = 0.80$ ($q = e_q^2 = 0.64$) and $e_q = 0.85$ ($q = 0.723$).

Figure~\ref{fig:4} shows that these two versions of calculations describe well the data of both experiments, and it is obvious that the normalization of the GT strength function for the experiment reported in~\cite{Thies_2012_100Mo} calls for a refinement in the region of energies of up to 20~MeV reckoned in the isotope $^{100}$Tc.

Thus, the calculations for the isotopes $^{98}$Mo and $^{100}$Mo have led to close values of the normalization of the GT strength functions or $\sum B_i(\mathrm{GT})$ (see~(\ref{eq:Summ.M_i})) -- from $q = e_q^2 = 0.64$ ($e_q = 0.80$) for $^{100}$Mo, to $q = 0.81$ ($e_q = 0.90$) for $^{98}$Mo isotope. This confirms the presence of the \textit{quenching}-effect.

A detailed analysis of the\textit{quenching}-effect was performed in the study~\cite{Lutostansky2022}, where it was found that $e_q = 0.90$ ($q = 0.81$) for the isotope $^{98}$Mo and $e_q = 0.8$ ($q = 0.64$) for the isotope $^{100}$Mo, which confirms the presence of the \textit{quenching}-effect.

\section{CROSS SECTIONS FOR SOLAR NEUTRINO CAPTURE BY $^{98,100}$Mo NUCLEI}

The $(\nu_e, e^{–})$ cross section, which depends on the incident-neutrino energy $E_{\nu}$, is
given by: 
\begin{equation} 
	\label{eq:sigma(E)}
	\begin{aligned}
		\sigma(E_{\nu}) = \frac{(G_F g_A)^2}{\pi c^3 \hbar^4} \int_{0}^{W - Q} W p_e F(Z, A, W) S(x) dx \\
		W = E_{\nu} - Q - x - m_e c^2 \\
		cp_e = \sqrt{W^{2} - (m_e c^2)^2}
	\end{aligned}
\end{equation}
where $F(Z, A, W)$ is the Fermi-function, $S(E)$ is the charge-exchange strength function, $G_F / (\hbar c)^3 = 1.1663787(6) \times 10^{-5}$~GeV$^{-2}$ is the Fermi weak coupling constant and $g_A = - 1.2723(23)$ is the axial-vector constant from \cite{PDG_2020}.

\begin{figure}[ht!]
	\centering
	\includegraphics[width=0.7\linewidth]{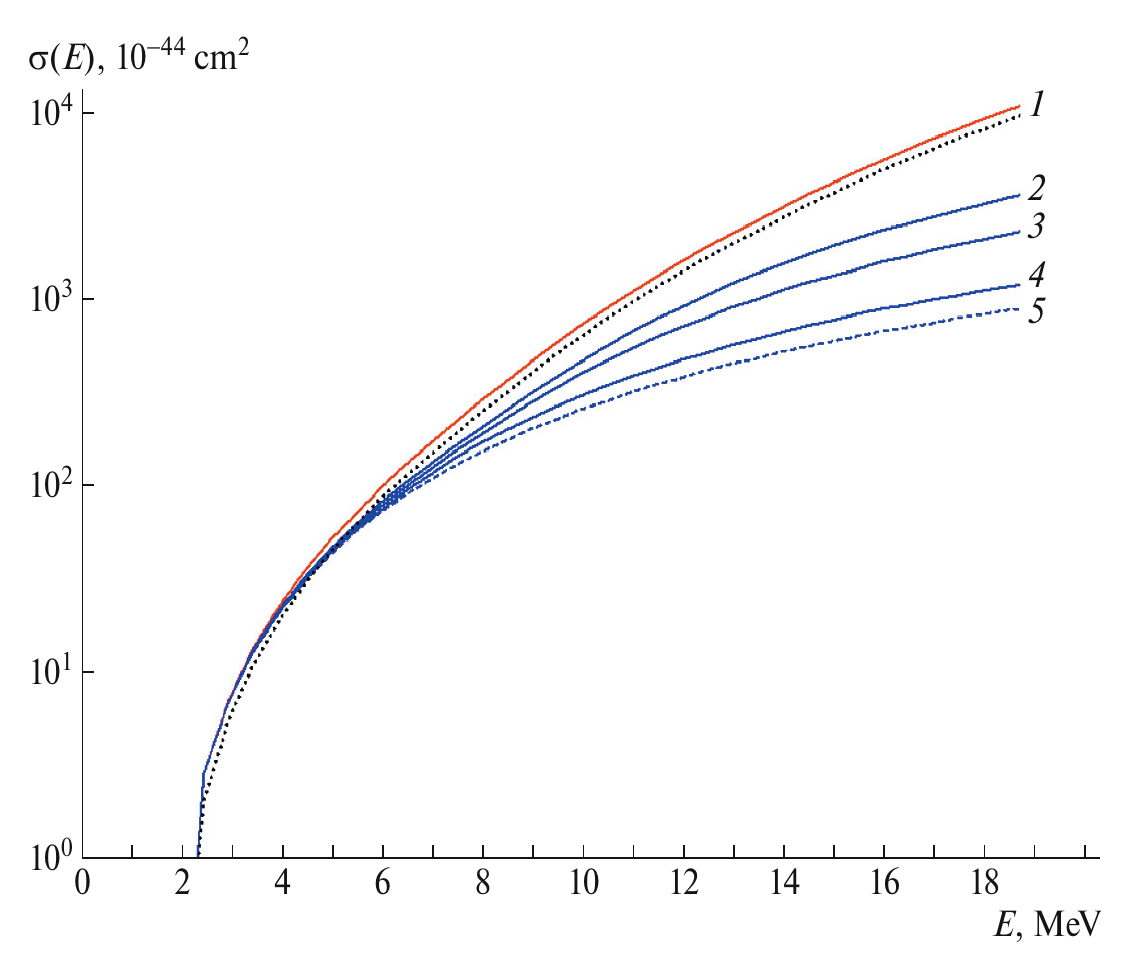}
	\caption{Neutrino-capture cross section $\sigma(E)$ in the reaction $^{98}\mathrm{Mo}(\nu_{e}, e^{-})^{98}\mathrm{Tc}$. 
		The points on display stand for the results of the calculation based on the experimental strength function $S(E)$ (see Fig.~\ref{fig:2}).
		 The solid and dashed curves represent the results of the calculations performed with the strength function $S(E)$ obtained within the TFFS approach: \textit{(1)} total cross section and (\textit{2}, \textit{3}, \textit{4}, and \textit{5})
		  results of the calculations not including, respectively, GTR; GTR and PR1; GTR, PR1, and PR2; and GTR, PR1, PR2, and PR3, where GTR is the GT resonance.
	  }
	\label{fig:5}
\end{figure}

\begin{figure}[ht!]
	\centering
	\includegraphics[width=0.7\linewidth]{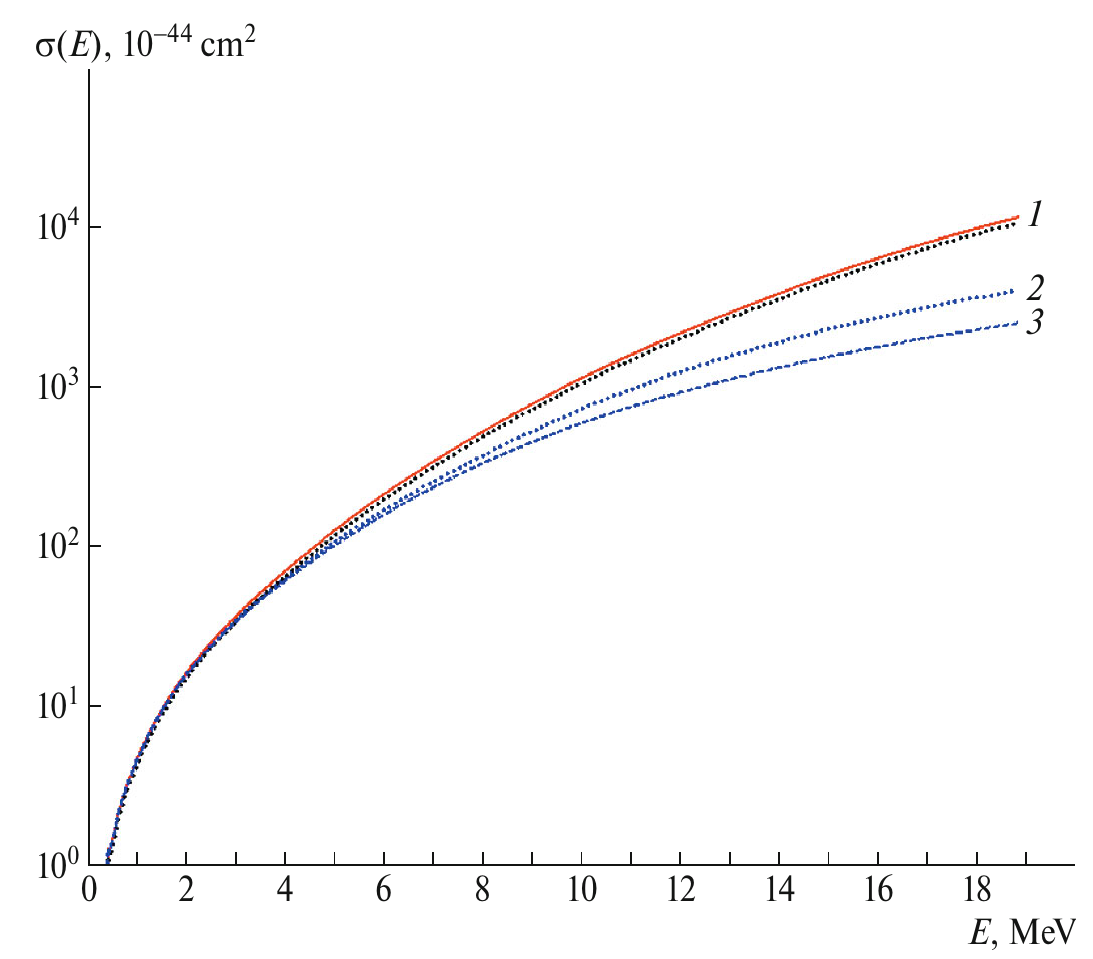}
	\caption{Neutrino-capture cross section $\sigma(E)$ in the reaction $^{100}\mathrm{Mo}(\nu_e, e^{-})^{100}\mathrm{Tc}$. The points on display 
	stand for the results of the calculation based on the experimental strength function $S(E)$ (see Fig.~\ref{fig:2}).
	The solid and dashed curves represent the results of the calculations performed with the strength
	function $S(E)$ obtained within the TFFS approach: \textit{(1)} total cross section, \textit{(2)} results obtained without including GTR, and \textit{(3)} results obtained without including GTR and PR1.}
	\label{fig:6}
\end{figure}

The neutrino-capture cross section $\sigma(E)$ is shown in Fig.~\ref{fig:5} for the reaction
$^{98}\mathrm{Mo}(\nu_e, e^{-})^{98}\mathrm{Tc}$ and in Fig.~\ref{fig:6} for the reaction $^{100}\mathrm{Mo}(\nu_e, e^{-})^{100}\mathrm{Tc}$. The cross sections
$\sigma(E)$ are given both according to the calculations with the experimental strength function $S(E)$ (see Fig.~\ref{fig:2}) and according to the calculations with the strength function
$S(E)$ obtained within the TFFS approach. Also, the results of the calculations performed without allowing for GT and pygmy resonances are presented. The figures show that the calculations with the strength functions $S(E$) obtained within the TFFS approach describe fairly well the cross sections $\sigma(E)$ calculated with the experimental strength functions, the average discrepancies for the total cross section not exceeding 10$\%$ both for $^{98}\mathrm{Mo}$ and for $^{100}\mathrm{Mo}$.

From Figs.~\ref{fig:5} and~\ref{fig:6}, one can see that the effect of charge-exchange resonances on the cross section $\sigma(E)$ is quite significant. 
The disregard of only two resonances, the GT resonance and PR1, reduces the cross section $\sigma(E)$ for $^{98}$Mo by a value of about 10$\%$ to a value of about 60$\%$ for neutrino energies between 4 and 14 MeV; for $^{100}$Mo,
the respective reduction is about 5$\%$ to 40$\%$. 
Thus, the effect of the resonances on the cross section $\sigma(E)$ for the $^{100}$Mo nucleus is smaller than for the $^{98}$Mo nucleus.

\begin{figure}[ht!]
	\centering
	\includegraphics[width=0.7\linewidth]{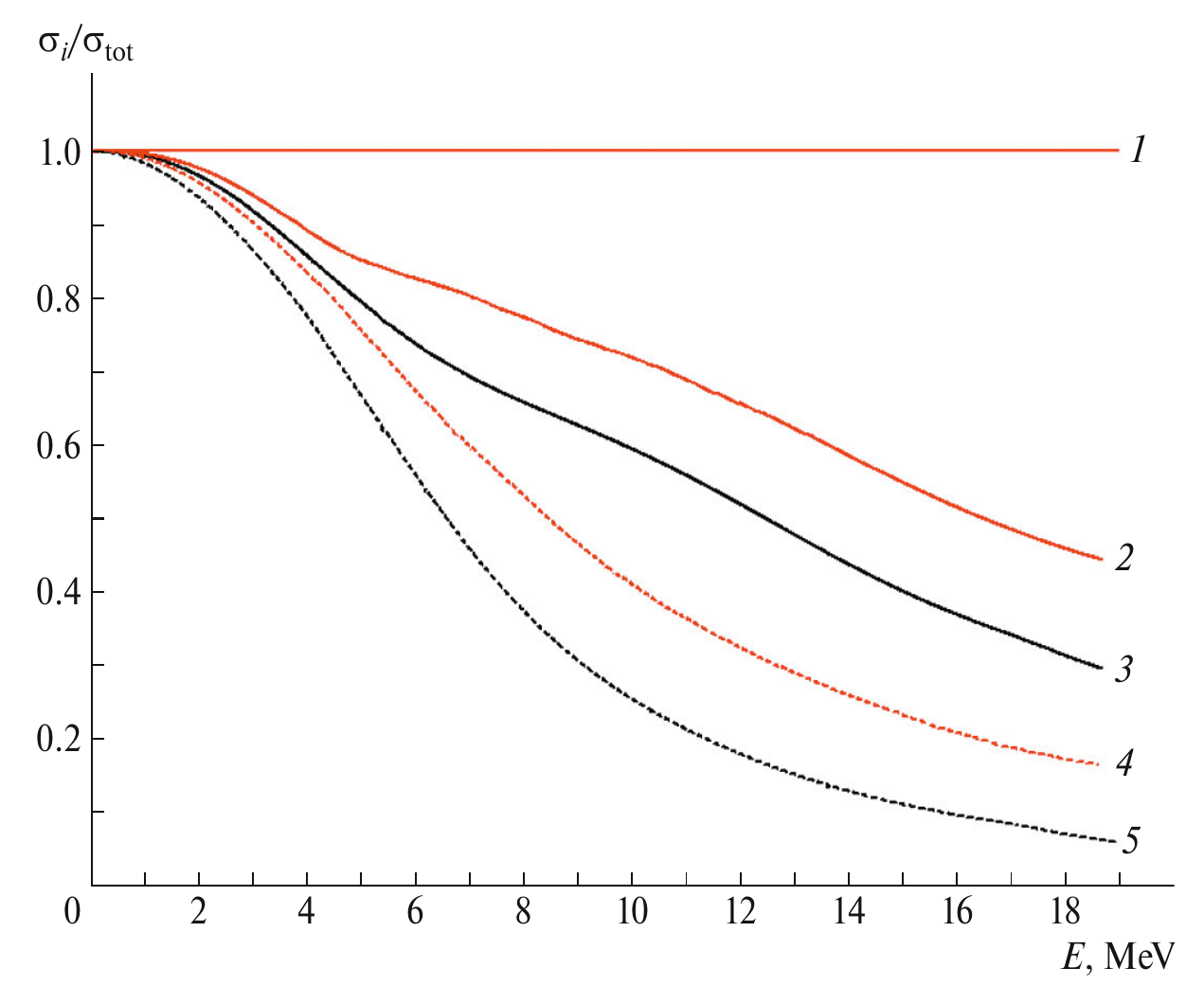}
	\caption{Ratios of the cross sections $\sigma_i(E)$ calculated for the reactions (curves \textit{3} and \textit{5})  $^{98}\mathrm{Mo}(\nu_e, e^{-})^{98}\mathrm{Tc}$ and (curves \textit{2} and \textit{4}) $^{100}\mathrm{Mo}(\nu_e, e^{-})^{100}\mathrm{Tc}$ to the total cross section $\sigma_{tot}(E)$ based on the TFFS approach (curve \textit{1}). Curves \textit{2} and \textit{3} were calculated without taking into account GTR, while curves
	\textit{4} and \textit{5} were calculated without taking into account GTR and PR1.}
	\label{fig:7}
\end{figure}

This can be seen in Fig.~\ref{fig:7}, where the relations of the calculated cross sections $\sigma_i(E)$ of
the reactions $^{98}\mathrm{Mo}(\nu_e, e^{-})^{98}\mathrm{Tc}$ and $^{100}\mathrm{Mo}(\nu_e, e^{-})^{100}\mathrm{Tc}$ are presented, normalized to the
total cross section $\sigma_{tot}(E)$ with the strength functions $S(E)$ calculated within the TFFS
framework. 
The reason behind the reduction of the effect of charge-exchange resonances on the cross section $\sigma(E)$ for neutrino capture by the $^{100}$Mo nucleus in
relation to the respective results for $^{98}$Mo is that the cross section for $^{100}$Mo receives a
contribution primarily from low-energy solar neutrinos, whose number is several
orders of magnitude greater than the number of neutrinos having energies in the 
region of $E_{\nu} > 2$~MeV and making a dominant contribution to the region of
resonances in $^{98}$Mo.

\section{RATE OF SOLAR-NEUTRINO CAPTURE BY $^{98,100}$Mo NUCLEI}

The solar-neutrino capture rate $R$ (number of neutrinos absorbed per unit time)
is related to the solar neutrino flux and the capture cross section by the equation:
\begin{equation} 
	\label{eq:R_total}
	R = \int_{0}^{E_{max}} \rho_{solar}(E_{\nu}) \sigma_{total}(E_{\nu}) dE_{\nu}
\end{equation}
where for the energy $E_{max}$, we can restrict ourselves (see [31]) to hep neutrinos (that
is, those from the reaction $^{3}\mathrm{He} + p \rightarrow {}^{4}\mathrm{He} + e^{+} + \nu_e)$, in which case $ E_{max} \le 18.79$
MeV, or to boron neutrinos (that is, those from the reaction ${}^8\mathrm{B} \rightarrow {}^{8}\mathrm{Be}^{*} + e^+ + \nu_e$, in
which case $ E_{max} \le 16.36 $ MeV. The solar-neutrino capture rate is given in SNU (SNU is a solar neutrino unit that corresponds to one detection event per second per $10^{36}$ target nuclei).

In calculating the cross sections for solar-neutrino capture, it is important to simulate correctly the flux of solar neutrinos. Several solar models have been vigorously developed in recent years. They include the BS05(OP), BS05(AGS, OP) and BS05(AGS, OPAL) models evolved by the group headed by J. N. Bahcall~\cite{Bahcall_2005}.
The helium concentration and metallicity (the specific number of atoms heavier than helium), as well as their distributions over the star volume, are the most important simulated parameters, along with the medium opaqueness parameter and the
dimensions of the convection zone. 
A description of neutrino fluxes requires a detailed knowledge of the cross section for neutrino interaction with a detector material and, as a consequence, knowledge of the strength function and its resonance structure for nuclei of this material. 
In this article, we present calculations based on the BS05(OP) model, which is the most convenient for a comparison with experimental data. 
The recalculations to other solar models are reduced to neutrino fluxes renormalization.

\begin{table}[ht]
	\caption{Solar-neutrino capture rate $R$ (in SNU) calculated for the isotope $^{98}$Mo with the strength function obtained from experimental data reported in~\cite{Rapaport_1985_98Mo} (the reduction (in percent) of the capture rates upon the disregard of GTR and GTR + PR1 is indicated parenthetically)}
	\centering
	\begin{adjustbox}{width=0.9\textwidth}
		\small
		\begin{tabular}{| c | c | c | c | c | c |} 
			\hline
			$^{98}$Mo & $^{8}$B & $hep$ & $^{15}$O & $^{17}$F & Total \\ 
			
			\hline
			$R$ & 18.415 & 0.105 & $5.1\cdot10^{-5}$ & $10^{-6}$ & 18.520 \\
			\hline
			
			$R$ without GTR & 10.893 (-40.8\%) & 0.044 (-58\%) & $5.1\cdot10^{-5}$ & $10^{-6}$ & 10.937 (-41\%) \\
			\hline
			
			$R$ without GTR and PR1 & 8.282 (-55\%) & 0.025 (-76\%) & $5.1\cdot10^{-5}$ & $10^{-6}$ & 8.307 (-55\%) \\
			\hline
			
		\end{tabular}
	\end{adjustbox}
	\label{table:1}
\end{table}

\begin{table}[ht]
	\caption{Solar-neutrino capture rate $R$ (in SNU) calculated for the isotope $^{98}$Mo with the strength function obtained within the TFFS framework~\cite{Lutostansky2022} (the reduction (in percent) of the capture rates upon the disregard of GTR and GTR + PR1 is indicated parenthetically)}
	\centering
	\begin{adjustbox}{width=0.9\textwidth}
		\small
		\begin{tabular}{| c | c | c | c | c | c |} 
			\hline
			$^{98}$Mo & $^{8}$B & $hep$ & $^{15}$O & $^{17}$F & Total \\ 
			
			\hline
			$R$ & 18.92 & 0.108 & 0.0002 & $10^{-5}$ & 19.028 \\
			\hline
			
			$R$ without GTR & 12.515 (-34\%) & 0.057 (-47\%) & 0.0002 & $10^{-5}$ & 12.572 (-34\%) \\
			\hline
			
			$R$ without GTR and PR1 & 10.778 (-43\%) & 0.043  (-60\%) & 0.0002 & $10^{-5}$ & 10.822 (-43\%) \\
			\hline
			
		\end{tabular}
	\end{adjustbox}
	\label{table:2}
\end{table}

\begin{table}[ht]
	\caption{Solar-neutrino capture rate $R$ (in~SNU) calculated for the isotope $^{100}$Mo with the strength function obtained from the experimental data reported in~\cite{AKIMUNE199723}, \cite{Thies_2012_100Mo} (also given here are the results of the calculations performed in~\cite{Ejiri-Elliot2014} and based on the use of the data from\cite{Thies_2012_100Mo}; the reduction (in percent) of the capture rates upon the disregard of GTR and GTR + PR1 is indicated parenthetically)}
	\centering
	\begin{adjustbox}{width=0.9\textwidth}
		\small
		\begin{tabular}{| c | c | c | c | c | c | c | c | c | c |} 
			\hline
			$^{100}$Mo & $pp$ & $pep$ & $^{7}$Be & $^{8}$B & $^{13}$N & $^{17}$F & $^{15}$O & $hep$ & Total \\ 
			
			\hline
			$R$ & 692.73 & 15.93 & 230.06 & 25.60 & 12.46 & 0.40 & 15.76 & 0.12 & 993.05\\
			
			\hline
			$R$ without GTR & 692.73 & 15.93 & 230.06 & 19.39
			(-24\%) & 12.46 & 0.40 & 15.76 & 0.07 (-42\%) & 986.78 (-0.6\%)\\
			
			\hline
			$R$ without GTR and PR1 & 692.73 & 15.93 & 230.06 & 15.82 (-38\%) & 12.46 & 0.40 & 15.76 & 0.05 (-58\%) & 983.21 (-1\%)\\	
			
			\hline	
			\cite{Ejiri-Elliot2014} & 695 & 16 & 234 & 16 & 12 &  & 16 &  & 989\\		
			\hline	
			
		\end{tabular}
	\end{adjustbox}
	\label{table:3}
\end{table}

\begin{table}[ht]
	\caption{Solar-neutrino capture rate $R$ (in~SNU) calculated for the isotope $^{100}$Mo with the strength function obtained within the TFFS framework~\cite{Lutostansky2022} (the reduction (in percent) of the capture rates upon the disregard of GTR and GTR + PR1 is indicated parenthetically)}
	\centering
	\begin{adjustbox}{width=0.9\textwidth}
		\small
		\begin{tabular}{| c | c | c | c | c | c | c | c | c | c |} 
			\hline
			$^{100}$Mo & $pp$ & $pep$ & $^{7}$Be & $^{8}$B & $^{13}$N & $^{17}$F & $^{15}$O & $hep$ & Total \\ 
			
			\hline
			$R$ & 586.58 & 14.46 & 202.16 & 31.42 & 10.91 & 0.35 & 14.08 & 0.15 & 860.11\\
			
			\hline
			$R$ without GTR & 586.58 & 14.35 & 201.80 & 20.61  (-34\%) & 10.89 & 0.35 & 14.01 & 0.08 (-47\%) & 848.67 (-1.3\%)\\
			
			\hline
			$R$ without GTR and PR1 & 586.58 & 14.29 & 201.61 & 17.28 (-45\%) & 10.88 & 0.35 & 13.97 & 0.06 (-60\%) & 845.02 (-1.8\%)\\		
			\hline	
			
		\end{tabular}
	\end{adjustbox}
	\label{table:4}
\end{table}

The numerical values of the solar-neutrino capture rates $R$ calculated for the
isotopes $^{98}$Mo and $^{100}$Mo are presented in Tables~\ref{table:1}--\ref{table:4}   (in~SNU). 
The tables give the results obtained by calculating $R$ with the experimental and theoretical strength functions $S(E)$ and with and without Gamow-Teller and Pygmy resonances [32]. 
The calculations with the experimental strength functions $S(E)$ (see Table~\ref{table:1} and~\ref{table:3}) were
performed by employing data obtained in the reactions $^{98}\mathrm{Mo}(p, n)^{98}\mathrm{Tc}$~\cite{Rapaport_1985_98Mo} and
$^{100}\mathrm{Mo}(^3\mathrm{He}, t)^{100}\mathrm{Tc}$~\cite{AKIMUNE199723}, \cite{Thies_2012_100Mo} (see Fig.~\ref{fig:2}).

The capture rate obtained for the isotope $^{98}$Mo is $R_{\mathrm{Total}} = 18.52$ SNU (Table~\ref{table:1}),
which is close to the value of SNU from~\cite{Bahcall_2005} and to the result found by employing the calculated strength functions; our result is $R_{\mathrm{Total}} = 19.028 $~SNU (Table~\ref{table:1}), while the respective result reported earlier in~\cite{Erokhina_1995} is $28^{+15}_{-8}$~SNU.

In the calculations for $^{100}$Mo (Table~\ref{table:3}), use was made of two sets of experimental data—one from~\cite{AKIMUNE199723} and the other from~\cite{Thies_2012_100Mo}. 
The point is that the table of the data presented~\cite{Thies_2012_100Mo} for the energies $E$ and matrix elements \textit{B}(GT) covers the energy range
of $ E \leq 4$~MeV, whereas, the article of H. Akimune and his coauthors~\cite{AKIMUNE199723}, which was published earlier, presents the tabulated data on high-lying excitations of the daughter nucleus $^{100}$Tc. 
In addition to the $R$ values calculated for $^{100}$Mo with the experimental strength functions, Table~\ref{table:3} gives the results obtained by H. Ejiri and S. R. Elliott~\cite{Ejiri-Elliot2014}
on the basis of the data from~\cite{Thies_2012_100Mo} extending to 4~MeV. 
In our results, this corresponds to the calculations that disregard GTR, and the discrepancies are insignificant,
whereas the discrepancies between the values of $R_{\mathrm{Total}}$ are approximately $0.4\%$. 
In the article published in 2017~\cite{Ejiri-Elliot2017}, the same authors presented the value of $R_{\mathrm{Total}} = 975$~SNU, which differs from their earlier value and from our estimate by about $1\%$. 
The discrepancies in question stem from special features of experimental data processing and are irrelevant to the present analysis.

Comparing the results of the calculations for $^{98}$Mo and $^{100}$Mo (Tables~\ref{table:1}, \ref{table:2} and \ref{table:3}, \ref{table:4}), first of all, it should be noted a large difference, more than 45 times, between the
values of $R_{\mathrm{Total}}$ for these isotopes. 
This is explained by a large difference between the
energy of $Q_1 = 1684$ keV for the isotope $^{98}$Tc and the energy of $Q_2 = 172.1$ keV for the isotope $^{100}$Tc (see Fig.~\ref{fig:1}). 
As a result, a dominant contribution to the process of
solar neutrinos capture comes from hard solar neutrinos in the case of the $^{98}$Mo nucleus and from lower energy neutrinos, mostly pp solar neutrinos, whose number is several orders of magnitude greater (see Fig.~\ref{fig:2}), in the case of the $^{100}$Mo nucleus.
For example, the contribution of hard boron neutrinos to $R_{\mathrm{Total}}$ in the case of $^{98}$Mo is $99\%$,
whereas their contribution in the case of $^{100}$Mo is as small as $2.6\%$; at the same time, soft pp neutrinos make a $70\%$ contribution in the latter case (see Fig.~\ref{fig:2}).

The discrepancies between the $R$ values obtained from the experimental and calculated data on the strength functions $S(E)$ are more significant, amounting, for
$R_{\mathrm{Total}}$ to about $3\%$ for $^{98}$Mo and to about $14\%$ for $^{100}$Mo. 
For $^{98}$Mo, this is explained by the discrepancies the description of resonance states~\cite{Lutostansky2022}, which make a dominant contribution to the neutrino capture cross section $\sigma(E_{\nu})$, and, for $^{100}$Mo, it is
explained by the inaccuracies in describing low-lying states, where the calculated value of $R$ depends greatly on the changes in $E_x$ and \textit{B}(GT). 
For example, the change in the ground-state position from 0 to 100~keV with a step of $\Delta E = 50 $ keV causes a
sequential change of about 150~SNU in $R_{\mathrm{Total}}$ at each step $\Delta E$ (in all, about 300~SNU).
Almost everything is associated here with the $pp$-neutrino channel. 
For neutrinos from $^7$Ве, the decrease is about 10~SNU at each step $\Delta E$.

The effect of charge-exchange resonances on the rates $R$ of solar neutrinos capture by the isotopes $^{98}$Mo and $^{100}$Mo is also illustrated in Tables~\ref{table:1}--\ref{table:4}. 
One can see that the values of $R_{\mathrm{Total}}$ for $^{100}$Mo undergo virtually no change in the case of the
calculation without GTR (decrease of about $1\%$) and in the case of the calculation without GTR and PR1 (decrease of about $2\%$), but that, for $^{98}$Mo, these changes are
significant: -$34\%$ and -$43\%$, respectively. 
As was indicated above, the reason is that a dominant contribution to $R_{\mathrm{Total}}$ comes from low-energy neutrinos (about $70\%$) -- mostly from $pp$ solar neutrinos -- for $^{100}$Mo and from boron solar neutrinos (about
$99\%$) for $^{98}$Mo. 
As a result, the calculations without GTR and PR1 make nearly
identical contributions to $R_{\mathrm{Total}}$ and $R({}^8\mathrm{B})$ for $^{98}\mathrm{Mo}$. The situation is similar for the iodine isotope $^{127}$I~\cite{LUTOSTANSKY-Physics-Letters-B}, in which case $R_{\mathrm{Total}} = 37.904$~SNU and $R({}^8\mathrm{B}) = 33.232$~SNU differ by a value as small as 12.3\%, whereas the GTR and PR1 contributions reduce $R_{\mathrm{Total}}$ by $72.7\%$ to 10.345 SNU ($27.3\%$) owing primarily mainly due to boron solar neutrinos.

The analog resonances at the energies of $E(\mathrm{AR})_{exp} = 9.7$~MeV~\cite{Rapaport_1985_98Mo} and $E(\mathrm{AR})_{calc} =
9.78$~MeV~\cite{Lutostansky2022} in $^{98}$Mo and at the energies of $E(\mathrm{AR})_{exp} = 11.085$~MeV~\cite{Thies_2012_100Mo} and $E(\mathrm{AR})_{calc} = 10.99$~MeV~\cite{Lutostansky2022} in $^(100)$Mo weakly affect the cross sections $\sigma(E)$ and on the solar-neutrino capture rates $R$. 
For example, the inclusion of the analog resonances increases $R$ by $\Delta R \leq 5\%$ for $^{98}$Mo and by $ \Delta R \leq 1\%$ for $^{100}$Mo.

\section{CONTRIBUTION OF BACKGROUND SOLAR NEUTRINOS
	IN DOUBLE BETA DECAY $^{100}$Mo}

According to the scheme of excited levels of the $^{100}$Mo nucleus (Fig.~\ref{fig:1}), all excitations up to the neutron separation energy $S_n = 6764.4$~keV will decay into the $^{100}$Tc ground state with subsequent decay into $^{100}$Ru. Such formation of the $^{100}$Ru nucleus will imitate the formation of this nucleus in the process of double beta decay and will be a background event in this process.

Without taking into account neutrino oscillations, our calculations of the number of neutrino events for  $^{100}$Mo up to the neutron-separation energy with allowance for both GT and analog resonances give a value of 188.37 events per ton per year. 
The calculations were performed up to the energy of neutron separation from the $^{100}$Tc nucleus, since higher lying excitations will be discharged via neutron emission and transitions to excited states of the $^{99}$Tc nucleus. Such process is outside the scope of this paper and should be examined separately. This will not contribute to backgrounds to $^{100}$Mo double-beta decay involving solar neutrinos. 
The articles published in recent years present the following values: in 2014 H. Ejiri and S. R. Elliott estimate $R = 989$~SNU~\cite{Ejiri-Elliot2014} (and $R = 975$~SNU~\cite{Ejiri-Elliot2017} later, in 2017), which, in terms of a ton of substance, gives 187.61~\cite{Ejiri-Elliot2014} (184.95~\cite{Ejiri-Elliot2017}) events per ton per year. 

\begin{table}[ht]
	\caption{Expected number of events per year (excluding oscillations) from solar neutrinos for current (with the exception of NEMO-3) and planned experiments to search for $0 \nu \beta \beta$ decay in $^{100}$Mo.}
	\centering
	\begin{adjustbox}{width=0.9\textwidth}
		\small
		\begin{tabular}{| c | c | c | c | c |} 
			\hline
			\multirow{2}{*}{\makecell{ Experiment}} & \multirow{2}{*}{\makecell{ Mass of target isotope \\ $^{100}$Mo, kg}} & \multicolumn{3}{|c|}{Expected number of events per year (without oscillation)} \\ 
			\cline{3-5}
			& & R = 993.05~\cite{Lutostansky2022_December} & R = 989~\cite{Ejiri-Elliot2014} & R = 975~\cite{Ejiri-Elliot2017} \\
			\hline
			$\mathbf{CUPID-Mo}$~\cite{Armengaud_cupid-mo} & 2.26 & 0.42 & 0.42 & 0.42 \\ 
			
			$\mathbf{CUPID}$~\cite{thecupidinterestgroup2019cupid} & Expected mass: 253 & 47.65 & 47.46 & 46.79 \\
			
			$\mathbf{AmoRE-I}$~\cite{AMORE-1} & 3.0 & 0.56 & 0.56 & 0.55 \\
			
			$\mathbf{AmoRE-II}$~\cite{AMORE-1} & Expected mass: 100  & 18.84 & 18.76 & 18.50 \\
			
			\multirow{2}{*}{\makecell{$\mathbf{CROSS}$~(medium scale \\ demonstrator)~\cite{CROSS-Bandac2020}}} & Expected mass: 4.7  & 0.88 & 0.88 & 0.87 \\
			& & & & \\

			$\mathbf{Mini-BINGO}$~\cite{BINGO-armatol2023new} & \multirow{2}{*}{\makecell{Expected mass (according to \\ our estimates): 3.36 }} & 0.63 & 0.63 & 0.62 \\
			& & & & \\
			
			\cite{CJPL-Chen2022} & \multirow{2}{*}{\makecell{Expected mass (according to \\ our estimates): 5.38 }} & 1.01 & 1.01 & 0.99 \\
			& & & & \\
			
			$\mathbf{NEMO-3}$~\cite{Arnold_NEMO3} & 6.91 & 1.30 & 1.30 & 1.28 \\
			\hline
			
		\end{tabular}
	\end{adjustbox}
	\label{table:5}
\end{table}

Table~\ref{table:5} presents the expected values of background events for actually operating (with the exception of NEMO-3) and planned experiments to search for neutrinoless double beta decay in the $^{100}$Mo isotope without taking into account oscillations. 
The CUPID-Mo experiment~\cite{Armengaud_cupid-mo} uses an array of $\mathrm{Li}_{2}\mathrm{MoO}_{4}$ crystals cooled to ~200~mK located in the underground laboratory in Modane (LSM) in France, in which the phonon and scintillation signals are detected. 
A similar approach was implemented in the AMoRE-I experiment~\cite{AMORE-1}, which uses the Yang Yang underground laboratory in South Korea ($\mathrm{Ca}\mathrm{MoO}_{4}$ and $\mathrm{Li}_{2}\mathrm{MoO}_{4}$ crystals). 
Particularly interesting, in the context of the search for the $0 \nu \beta \beta$ process, are the continuations of these experiments: CUPID~\cite{thecupidinterestgroup2019cupid} and AMoRE-II~\cite{AMORE-1} because in them, the planned mass of the $^{100}$Mo isotope will exceed hundreds of kilograms, which will make it possible to achieve an exposure of more than tons*year.
New technologies for detecting the $0 \nu \beta \beta$ process are being developed in the CROSS~\cite{CROSS-Bandac2020}, BINGO~\cite{BINGO-armatol2023new} collaborations and in a planned experiment at the CJPL underground laboratory in China~\cite{CJPL-Chen2022}.

Using the data and reports of the NEMO-3 experiment as an example, let us analyze the ratio of the number of background events from different sources. 
Thus, according to~\cite{Arnold-Nemo3-2015} (Table 8), during an exposure of 34.3 kg$\cdot$years, 15 background two-electron events were observed in the energy range sought for in the experiment energy range from 2.8 to 3.2~MeV. According to our estimate, for the same parameters, the contribution of background events from the absorption of solar neutrinos will be about $\le 1\%$, i.e. $\le 0.15$ events, excluding oscillations. 
Accounting for this effect will reduce the number of events by approximately half. It should also be noted that, depending on the design features of the experiment, the background from solar neutrinos can be significantly suppressed~\cite{Ejiri_2016}.

\section{CONCLUSIONS}

We have explored the interaction of solar neutrinos with $^{98}$Mo and $^{100}$Mo nuclei, taking into account the effect of charge-exchange resonances. We have studied the effect of high-lying charge-exchange resonances in the strength function $S(E)$ on the cross sections for solar-neutrino capture by $^{98}$Mo and $^{100}$Mo nuclei. We have employed both experimental data obtained for the strength functions $S(E)$ in $(p, n)$ and $(^{3}\mathrm{He}, t)$ charge-exchange reactions~\cite{Rapaport_1985_98Mo}, \cite{AKIMUNE199723}, \cite{Thies_2012_100Mo} and the strength functions $S(E)$ calculated within the theory of finite Fermi-systems~\cite{Lutostansky2022}.

A comparison of the calculated strength function $S(E)$ with experimental data shows good agreement both in resonance-peak energy and in resonance peak amplitude. 
There is a deficit in the sum rule for GT excitations. It is due either to the \textit{quenching}-effect~\cite{ARIMA1999260} or to a violation of the normalization of GT matrix elements. 
Within the TFFS framework~\cite{Migdal_book}, this deficit is compensated by the introduction of an effective charge $e_q = 0.90$ $(q = 0.81)$ for the isotope $^{98}$Mo and $e_q = 0.8$ $(q = 0.64)$ for the isotope $^{100}$Mo~\cite{Lutostansky2022}.

We have calculated the cross sections $\sigma(E)$ for solar-neutrino capture and have analyzed the contribution of all charge-exchange resonances. We have found that, in all energy ranges, the cross section $\sigma(E)$ is substantially larger for the $^{100}$Mo nucleus
than for the $^{98}$Mo nucleus. 
This is because the energy thresholds $Q_1$ and $Q_2$ are markedly different for the neighboring isobaric nuclei $^{98}$Tc and $^{100}$Tc (see Fig.~\ref{fig:1}). 
As a result, the cross sections $\sigma(E)$ for the $^{98}$Mo and $^{100}$Mo nuclei are also different. 
Thus, low energy $pp$ neutrinos, whose number is several orders of magnitude greater, make the main contribution to $\sigma(E)$ $^{100}$Mo (see Fig.~\ref{fig:2}), while the resonant energy region does not affect. 
Accordingly, the contribution of high-energy nuclear resonances to $\sigma(E)$ is smaller for $^{100}$Mo than for $^{98}$Mo.

We have also calculated solar-neutrino capture rates $R$ for the isotopes $^{98}$Mo and $^{100}$Mo, taking into account all components of the solar neutrino spectrum. 
The calculations have been performed both with experimental and with theoretical strength functions $S(E)$ and with and without allowance for the GT and pygmy resonances.

Comparing the results of our calculations for $^{98}$Mo and $^{100}$Mo nuclei, we note that the values of $R_{\mathrm{Total}}$ for these isotopes differ by a factor greater than 45. 
This is explained by the fact that, as it was already indicated, in the process of capturing solar neutrinos by the $^{98}$Mo nucleus, the main role is played by the high energy solar neutrinos, and in the case of the $^{100}$Mo nucleus - by neutrinos with lower energies, which are orders of magnitude greater.

Thus, the two isotopes $^{98}$Mo and $^{100}$Mo of the same element, which differ only slightly in structure and in charge-exchange strength function, differ sharply in solar-neutrino capture cross section, $\sigma(E)$, and in solar-neutrino capture rate.

The contribution of background solar neutrinos to the double-beta decay of $^{100}$Mo nuclei is estimated. It is shown that the source of background events, which is about 1$\%$ of the required statistics, ceases to be a minor background for exposures of the order of tons*year, and cannot be ignored during processing and additional data collection. As we showed in this article, the irremovable background from the capture of solar neutrinos by Mo nuclei is estimated at this level and requires either an increase in statistics to guarantee the result, or the use of experimental schemes to exclude events of this kind.

\section{ACKNOWLEDGMENTS}
We are are grateful to M. D. Skorokhvatov, I. N. Borzov, L. V. Inzhechik, V. V. Khrushchev S. S. Semenov, and A. K. Vyborov for stimulating discussions and for their help in work.

\section{FUNDING}
This work was supported in part by Russian Science Foundation (project no. 21-12-00061) and by a grant from the Department of Neutrino Processes at
National Research Center Kurchatov Institute.

\printbibliography

@book{Bahcall_book,
	author         = "J. N. Bahcall",
	title          = {{Neutrino Astrophysics}},
	publisher        = "Cambridge University Press. 1988.",
}

@article{AKIMUNE199723,
	journal = {Physics Letters B},
	volume = {394},
	number = {1},
	pages = {23-28},
	year = {1997},
	issn = {0370-2693},
	doi = {https://doi.org/10.1016/S0370-2693(96)01659-0},
	url = {https://www.sciencedirect.com/science/article/pii/S0370269396016590},
	author = {Hidetoshi Akimune and Hiroyasu Ejiri and Mamoru Fujiwara and Izuru Daito and Toru Inomata and Ryota Hazama and Atsushi Tamii and Hidenori Toyokawa and Mosaru Yosoi},
}

@article{Rapaport_1985_98Mo,
	author = {Rapaport, J. and Welch, P. and Bahcall, J. and Sugarbaker, E. and Taddeucci, T. N. and Goodman, C. D. and Foster, C. F. and Horen, D. and Gaarde, C. and Larsen, J. and Masterson, T.},
	journal = {Phys. Rev. Lett.},
	volume = {54},
	issue = {21},
	pages = {2325--2328},
	numpages = {0},
	year = {1985},
	month = {May},
	publisher = {American Physical Society},
	doi = {10.1103/PhysRevLett.54.2325},
	url = {https://link.aps.org/doi/10.1103/PhysRevLett.54.2325}
}

@article{Thies_2012_100Mo,
	author = {Thies, J. H. and Adachi, T. and Dozono, M. and Ejiri, H. and Frekers, D. and Fujita, H. and Fujita, Y. and Fujiwara, M. and Grewe, E.-W. and Hatanaka, K. and Heinrichs, P. and Ishikawa, D. and Khai, N. T. and Lennarz, A. and Matsubara, H. and Okamura, H. and Oo, Y. Y. and Puppe, P. and Ruhe, T. and Suda, K. and Tamii, A. and Yoshida, H. P. and Zegers, R. G. T.},
	journal = {Phys. Rev. C},
	volume = {86},
	issue = {4},
	pages = {044309},
	numpages = {8},
	year = {2012},
	month = {Oct},
	publisher = {American Physical Society},
	doi = {10.1103/PhysRevC.86.044309},
	url = {https://link.aps.org/doi/10.1103/PhysRevC.86.044309}
}

@article{Arnold_NEMO3,
	author = {Arnold, R. and Augier, C. and Barabash, A. S. and Basharina-Freshville, A. and Blondel, S. and Blot, S. and Bongrand, M. and Boursette, D. and Brudanin, V. and Busto, J. and Caffrey, A. J. and Calvez, S. and Cascella, M. and Cerna, C. and Cesar, J. P. and Chapon, A. and Chauveau, E. and Chopra, A. and others},
	DOI= "10.1140/epjc/s10052-019-6948-4",
	url= "https://doi.org/10.1140/epjc/s10052-019-6948-4",
	journal = {Eur. Phys. J. C},
	year = 2019,
	volume = 79,
	number = 5,
	pages = "440",
}

@article{RakhimovBarabash,
	url = {https://doi.org/10.1515/ract-2019-3129},
	author = {Alimardon V. Rakhimov and A. S. Barabash and A. Basharina-Freshville and S. Blot and M. Bongrand and Ch. Bourgeois and D. Breton and R. Breier and E. Birdsall and V. B. Brudanin and H. Burešova and J. Busto and S. Calvez and M. Cascella and C. Cerna and J. P. Cesar and E. Chauveau and A. Chopra and G. Claverie and S. De Capua and others},
	pages = {87--97},
	volume = {108},
	number = {2},
	journal = {Radiochimica Acta},
	doi = {doi:10.1515/ract-2019-3129},
	year = {2020},
	lastchecked = {2023-03-29}
}

@article{Armengaud_cupid-mo,
	author = {Armengaud, E. and Augier, C. and Barabash, A. S. and Bellini, F. and Benato, G. and Benoit, A. and Beretta, M. and Berge, L. and others},
	collaboration = {CUPID-Mo Collaboration},
	journal = {Phys. Rev. Lett.},
	volume = {126},
	issue = {18},
	pages = {181802},
	numpages = {7},
	year = {2021},
	month = {May},
	publisher = {American Physical Society},
	doi = {10.1103/PhysRevLett.126.181802},
	url = {https://link.aps.org/doi/10.1103/PhysRevLett.126.181802}
}

@Article{CUPID-Alfonso2022,
	author={Alfonso, K.
	and Armatol, A.
	and Augier, C.
	and Avignone, F. T.
	and Azzolini, O.
	and Balata, M.
	and Barabash, A. S.
	and others},
	journal={Journal of Low Temperature Physics},
	year={2022},
	month={Nov},
	day={29},
	issn={1573-7357},
	doi={10.1007/s10909-022-02909-3},
	url={https://doi.org/10.1007/s10909-022-02909-3}
}

@Article{ Augier_cupid-mo,
	author = {Augier, C. and Barabash, A. S. and Bellini, F. and Benato, G. and Beretta, M. and Berg\'e, L. and Billard, J. and Borovlev, Yu. A. and Cardani, L. and Casali, N. and Cazes, A. and Chapellier, M. and Chiesa, D. and Dafinei, I. and Danevich, F. A. and De Jesus, M. and de Marcillac, P. and Dixon, T. and others},
	DOI= {10.1140/epjc/s10052-022-10942-5},
	url= {https://doi.org/10.1140/epjc/s10052-022-10942-5},
	journal = {Eur. Phys. J. C},
	year = {2022},
	volume = {82},
	number = {11},
	pages = {1033},
}

@article{AMORE-1,
	author = {Kim, Hameed and HA, Dae and Jeon, E. and Jeon, Juyeal and Jo, H. and Kang, C. and Kang, W. and Kim, H. and Kim, Sandra and Kim, S. and Kim, Sun and Kim, S. and Kim, W. and Kim, Young-Seog and Kim, Young Ho and Kwon, D. and Lee, E. and Lee, Hae June and Lee, Hahn and Yoon, Young},
	year = {2022},
	month = {10},
	pages = {1-9},
	volume = {209},
	journal = {Journal of Low Temperature Physics},
	doi = {10.1007/s10909-022-02880-z}
}

@article{AMORE_Lee_2020,
	doi = {10.1088/1748-0221/15/08/C08010},
	url = {https://dx.doi.org/10.1088/1748-0221/15/08/C08010},
	year = {2020},
	month = {aug},
	publisher = {},
	volume = {15},
	number = {08},
	pages = {C08010},
	author = {M.H. Lee},
	journal = {Journal of Instrumentation},
}

@article{ Nesterenko-98Mo,
	author = {Nesterenko, D. A. and Jokiniemi, L. and Kotila, J. and Kankainen, A. and Ge, Z. and Eronen, T. and Rinta-Antila, S. and Suhonen, J.},
	DOI= "10.1140/epja/s10050-022-00695-w",
	url= "https://doi.org/10.1140/epja/s10050-022-00695-w",
	journal = {Eur. Phys. J. A},
	year = 2022,
	volume = 58,
	number = 3,
	pages = "44",
}

@article{Gaponov-Lutostansky-JETP-Lett,
	author = "given=Yuri, given-i={Yu. V}, family=Gaponov and given=Yuri, given-i={Yu. S}, family=Lutostansky,",
	journal = "JETP Lett.",
	volume = "15",
	pages = "120",
	year = "1972"
}

@article{Gaponov-Lutostansky-Nucl-Phys,
	author = "given=Yuri, given-i={Yu. V}, family=Gaponov and given=Yuri, given-i={Yu. S}, family=Lutostansky,",
	journal = "Sov. J. Nucl. Phys.",
	volume = "16",
	pages = "270",
	year = "1972"
}

@article{Lutostansky2017_JETP,
	author={given=Yuri, given-i={Yu. S}, family=Lutostansky, },
	title={Charge-exchange pigmy resonances of tin isotopes},
	journal={JETP Letters},
	year={2017},
	day={01},
	volume={106},
	number={1},
	pages={7-11},
	issn={1090-6487},
	doi={10.1134/S0021364017130112},
	url={https://doi.org/10.1134/S0021364017130112}
}

@article{Pham-PhysRevC,
	author = {Pham, K. and J\"anecke, J. and Roberts, D. A. and Harakeh, M. N. and Berg, G. P. A. and Chang, S. and Liu, J. and Stephenson, E. J. and Davis, B. F. and Akimune, H. and Fujiwara, M.},
	journal = {Phys. Rev. C},
	volume = {51},
	issue = {2},
	pages = {526--540},
	numpages = {0},
	year = {1995},
	month = {Feb},
	publisher = {American Physical Society},
	doi = {10.1103/PhysRevC.51.526},
	url = {https://link.aps.org/doi/10.1103/PhysRevC.51.526}
}

@article{Lutostansky_Tikhonov2018,
	%author={Yu. S. Lutostansky, and Tikhonov, V. N.},
	author={given=Yuri, given-i={Yu. S}, family=Lutostansky and Tikhonov, V. N.},
	journal={Physics of Atomic Nuclei},
	year={2018},
	day={01},
	volume={81},
	number={5},
	pages={540-549},
	issn={1562-692X},
	doi={10.1134/S1063778818040117},
	url={https://doi.org/10.1134/S1063778818040117}
}

@article{Verney-PhysRevC.95.054320,
	author = {Verney, D. and Testov, D. and Ibrahim, F. and Penionzhkevich, Yu. and Roussi\`ere, B. and Smirnov, V. and Didierjean, F. and Flanagan, K. and Franchoo, S. and Kuznetsova, E. and Li, R. and Marsh, B. and Matea, I. and Pai, H. and Sokol, E. and Stefan, I. and Suzuki, D.},
	journal = {Phys. Rev. C},
	volume = {95},
	issue = {5},
	pages = {054320},
	numpages = {14},
	year = {2017},
	month = {May},
	publisher = {American Physical Society},
	doi = {10.1103/PhysRevC.95.054320},
	url = {https://link.aps.org/doi/10.1103/PhysRevC.95.054320}
}

@article{Lutostansky:2019iri,
	author = "given=Yuri, given-i={Yu. S}, family=Lutostansky,  and Osipenko, A.P. and Tikhonov, V.N.",
	doi = "10.3103/S1062873819040178",
	journal = "Bull. Russ. Acad. Sci. Phys.",
	volume = "83",
	number = "4",
	pages = "488--492",
	year = "2019"
}

@article{Huang_2021,
	doi = {10.1088/1674-1137/abddb0},
	url = {https://dx.doi.org/10.1088/1674-1137/abddb0},
	year = {2021},
	month = {mar},
	publisher = {Chinese Physical Society and the Institute of High Energy Physics of the Chinese Academy of Sciences and the Institute of Modern Physics of the Chinese Academy of Sciences and IOP Publishing Ltd},
	volume = {45},
	number = {3},
	pages = {030002},
	author = {W.J. Huang and Meng Wang and F.G. Kondev and G. Audi and S. Naimi},
	journal = {Chinese Physics C},
}

@Article{Lutostansky2022,
	author={given=Yuri, given-i={Yu. S}, family=Lutostansky
	and Koroteev, G. A.
	and given=A., given-i={A. Yu.}, family=Lutostansky
	and Osipenko, A. P.
	and Tikhonov, V. N.
	and Fazliakhmetov, A. N.},
	journal={Physics of Atomic Nuclei},
	year={2022},
	month={Jun},
	day={01},
	volume={85},
	number={3},
	pages={231-240},
	issn={1562-692X},
	doi={10.1134/S1063778822030127},
	url={https://doi.org/10.1134/S1063778822030127}
}

@book{Migdal_book,
	author         = "A. B. Migdal,",
	title          = {{Theory of Finite Fermi Systems and Applications to Atomic Nuclei}},
	publisher        = "Nauka, Moscow, 1983, 2nd ed.; Interscience, New, York, 1967, transl. 1st ed.",
}

@article{Lutostansky2018_EPJ,
	author = {given=Yuri, given-i={Yu. S}, family=Lutostansky, },
	DOI= "10.1051/epjconf/201819402009",
	url= "https://doi.org/10.1051/epjconf/201819402009",
	journal = {EPJ Web Conf.},
	year = 2018,
	volume = 194,
	pages = "02009",
}

@article{LUTOSTANSKY-Physics-Letters-B,
	journal = {Physics Letters B},
	volume = {826},
	pages = {136905},
	year = {2022},
	issn = {0370-2693},
	doi = {https://doi.org/10.1016/j.physletb.2022.136905},
	url = {https://www.sciencedirect.com/science/article/pii/S0370269322000399},
	author = {given=Yuri, given-i={Yu. S}, family=Lutostansky and Almaz N. Fazliakhmetov and Grigory A. Koroteev and Nadezhda V. Klochkova and given=Andrey, given-i={A. Yu.}, family=Lutostansky and Alexey P. Osipenko and Victor N. Tikhonov},
}

@article{BORZOV1995335,
	journal = {Nuclear Physics A},
	volume = {584},
	number = {2},
	pages = {335-361},
	year = {1995},
	issn = {0375-9474},
	doi = {https://doi.org/10.1016/0375-9474(94)00769-J},
	url = {https://www.sciencedirect.com/science/article/pii/037594749400769J},
	author = {I.N. Borzov and S.A. Fayans and E.L. Trykov},
}

@article{Lutostansky2020,
	author = {given=Yuri, given-i={Yu. S}, family=Lutostansky, },
	year = {2020},
	pages = {33-38},
	volume = {83},
	journal = {Physics of Atomic Nuclei},
	doi = {10.1134/S106377882001007X}
}

@article{Lutostansky_Shulgina_PhysRevLett.67.430,
	author = { given=Yuri, given-i={Yu. S}, family=Lutostansky,  and Shul'gina, N. B.},
	journal = {Phys. Rev. Lett.},
	volume = {67},
	issue = {4},
	pages = {430--432},
	numpages = {0},
	year = {1991},
	publisher = {American Physical Society},
	doi = {10.1103/PhysRevLett.67.430},
	url = {https://link.aps.org/doi/10.1103/PhysRevLett.67.430}
}

@article{ARIMA1999260,
	journal = "Nuclear Physics A",
	volume = "649",
	number = "1",
	pages = "260 - 270",
	year = "1999",
	note = "Giant Resonances",
	issn = "0375-9474",
	doi = "https://doi.org/10.1016/S0375-9474(99)00070-6",
	url = "http://www.sciencedirect.com/science/article/pii/S0375947499000706",
	author = "A. Arima",
}

@article{Migdal-1957,
	author = "A. B. Migdal",
	journal = "Sov. Phys. JETP",
	volume = "5",
	pages = "333",
	year = "1957"
}

@article{Ejiri-Elliot2014,
	author = {Ejiri, H. and Elliott, S. R.},
	journal = {Phys. Rev. C},
	volume = {89},
	issue = {5},
	pages = {055501},
	numpages = {7},
	year = {2014},
	month = {May},
	publisher = {American Physical Society},
	doi = {10.1103/PhysRevC.89.055501},
	url = {https://link.aps.org/doi/10.1103/PhysRevC.89.055501}
}

@article{Ejiri-Elliot2017,
	author = {Ejiri, H. and Elliott, S. R.},
	journal = {Phys. Rev. C},
	volume = {95},
	issue = {5},
	pages = {055501},
	numpages = {5},
	year = {2017},
	month = {May},
	publisher = {American Physical Society},
	doi = {10.1103/PhysRevC.95.055501},
	url = {https://link.aps.org/doi/10.1103/PhysRevC.95.055501}
}

@Article{Lutostansky2022_December,
	author={given=Yuri, given-i={Yu. S}, family=Lutostansky
	and Belogortseva, N. A.
	and Koroteev, G. A.
	and given=A., given-i={A. Yu.}, family=Lutostansky
	and Osipenko, A. P.
	and Tikhonov, V. N.
	and Fazliakhmetov, A. N.},
	journal={Physics of Atomic Nuclei},
	year={2022},
	month={Dec},
	day={01},
	volume={85},
	number={6},
	pages={551-560},
	issn={1562-692X},
	doi={10.1134/S1063778822060096},
	url={https://doi.org/10.1134/S1063778822060096}
}

@article{thecupidinterestgroup2019cupid,
	author={The CUPID Interest Group},
	title={CUPID pre-CDR}, 
	year={2019},
	eprint={1907.09376},
	archivePrefix={arXiv},
	primaryClass={physics.ins-det},
	url = {
	https://doi.org/10.48550/arXiv.1907.09376},
}

@Article{CROSS-Bandac2020,
	author={Bandac, I. C.
	and Barabash, A. S.
	and Berg{\'e}, L.
	and Bri{\`e}re, M.
	and Bourgeois, C.
	and Carniti, P.
	and Chapellier, M.
	and de Combarieu, M.
	and Dafinei, I.
	and Danevich, F. A.
	and Dosme, N.
	and Doullet, D.
	and Dumoulin, L.
	and Ferri, F.
	and Giuliani, A.
	and Gotti, C.
	and others},
	journal={Journal of High Energy Physics},
	year={2020},
	month={Jan},
	day={07},
	volume={2020},
	number={1},
	pages={18},
	issn={1029-8479},
	doi={10.1007/JHEP01(2020)018},
	url={https://doi.org/10.1007/JHEP01(2020)018}
}

@article{BINGO-armatol2023new,
title={New results about the revolutionary bolometer assembly of BINGO}, 
author={A. Armatol},
year={2023},
eprint={2301.06946},
archivePrefix={arXiv},
primaryClass={physics.ins-det},
url = {https://doi.org/10.48550/arXiv.2301.06946}
}

@Article{CJPL-Chen2022,
	author={Chen, W.
	and Ma, L.
	and Chen, J. H.
	and Huang, H. Z.
	and Ma, Y. G.},
	journal={The European Physical Journal C},
	year={2022},
	month={Jun},
	day={22},
	volume={82},
	number={6},
	pages={549},
	issn={1434-6052},
	doi={10.1140/epjc/s10052-022-10501-y},
	url={https://doi.org/10.1140/epjc/s10052-022-10501-y}
}

@article{Ejiri_2016,
	doi = {10.1088/0954-3899/43/4/045201},
	url = {https://dx.doi.org/10.1088/0954-3899/43/4/045201},
	year = {2016},
	month = {feb},
	publisher = {IOP Publishing},
	volume = {43},
	number = {4},
	pages = {045201},
	author = {Hiroyasu Ejiri and Kai Zuber},
	journal = {Journal of Physics G: Nuclear and Particle Physics},
}

@article{PDG_2020,
	author = { Zyla, P A and Barnett, R M and Beringer, J and others},
	journal = {Progress of Theoretical and Experimental Physics},
	volume = {2020},
	number = {8},
	year = {2020},
	month = {08},
	issn = {2050-3911},
	%    doi = {10.1093/ptep/ptaa104},
	url = {https://doi.org/10.1093/ptep/ptaa104},
	note = {083C01},
	eprint = {https://academic.oup.com/ptep/article-pdf/2020/8/083C01/34673722/ptaa104.pdf},
}

@article{Bahcall_2005,
	doi = {10.1086/428929},
	url = {https://doi.org/10.1086/428929},
	year = 2005,
	publisher = {American Astronomical Society},
	volume = {621},
	number = {1},
	pages = {L85--L88},
	author = {John N. Bahcall and Aldo M. Serenelli and Sarbani Basu},
	journal = {The Astrophysical Journal}
}

@article{Erokhina_1995,
	doi = {10.1088/0031-8949/1995/T56/043},
	url = {https://dx.doi.org/10.1088/0031-8949/1995/T56/043},
	year = {1995},
	month = {jan},
	publisher = {},
	volume = {1995},
	number = {T56},
	pages = {258},
	author = {K I Erokhina and  V I Isakov},
	journal = {Physica Scripta},
}

@article{Arnold-Nemo3-2015,
	author = {Arnold, R. and Augier, C. and Baker, J. D. and Barabash, A. S. and Basharina-Freshville, A. and Blondel, S. and Blot, S. and Bongrand, M. and Brudanin, V. and Busto, J. and Caffrey, A. J. and Calvez, S. and Cerna, C. and Cesar, J. P. and Chapon, A. and others},
	collaboration = {NEMO-3 Collaboration},
	journal = {Phys. Rev. D},
	volume = {92},
	issue = {7},
	pages = {072011},
	numpages = {23},
	year = {2015},
	month = {Oct},
	publisher = {American Physical Society},
	doi = {10.1103/PhysRevD.92.072011},
	url = {https://link.aps.org/doi/10.1103/PhysRevD.92.072011}
}

\end{document}